\documentstyle[12pt,aaspp4]{article}
\newcommand\snra{G003.8--00.3}
\newcommand\snrb{G350.0--02.0}
\newcommand\etal{{\it et~al. }}
\newcommand\HI{H\,{\sc i}}

\begin{document}
\title{The nature of bilateral supernova remnants}
\author{B. M. Gaensler\altaffilmark{1,2}}
\altaffiltext{1}{Australia Telescope National Facility, CSIRO, PO Box 76,
Epping, NSW 2121, Australia}
\altaffiltext{2}{Astrophysics Department, School of Physics A29,
University of Sydney, NSW 2006, Australia;
b.gaensler@physics.usyd.edu.au}

\begin{center}
\today
\vspace{1cm}

This paper has been accepted to {\em The Astrophysical Journal}. 
\end{center}

\begin{abstract} We present high-resolution radio images at 1.4~GHz of
two Galactic supernova remnants (SNRs), \snra\ (formerly G003.7--00.2)
and \snrb\ (formerly G350.0-01.8).  Although the two objects are very
different in appearance, in both cases the radio emission shows a clear
bilateral (or ``barrel'') morphology for which the axis is parallel to
the Galactic Plane.

The majority of Galactic SNRs have now been observed at high
resolution, and one can define a clear bilateral subset of the
population.  We consider a sample of 17 such SNRs, and find a highly
significant tendency for the bilateral axes of these SNRs to be aligned
with the Galactic Plane.  We interpret this as indicating that
``extrinsic'' effects dominate the morphology of such remnants.
Specifically, we argue that the Galactic magnetic field causes these
SNRs to appear bilateral, either directly, in the form of magnetic
field compression and/or quasi-perpendicular acceleration of electrons
in the supernova shock, or more likely indirectly, by pre-processing
the interstellar medium to produce density stratifications extended
along the Plane.  

\end{abstract}


\clearpage

\section{Introduction}
\label{sec_intro}

High resolution observations of radio supernova remnants
(SNRs) have revealed exceedingly complex structures. The
appearance of a SNR can tell us much about the supernova (SN) explosion
itself, the nature of any central compact object, the circumstellar
medium (CSM) of the progenitor, the surrounding interstellar medium
(ISM), and the properties of the ambient magnetic field. Indeed the
difficulty in interpreting SNR morphologies has always been in
disentangling the competing effects and in determining
which ones dominate during the different evolutionary stages.

Perhaps the longest running debate concerning SNR morphologies has been
over the classification referred to as bilateral, bipolar, axially
symmetric, or ``barrel'' SNRs. These remnants are characterised by a
clear axis of symmetry, low levels of emission along this axis,
and two bright limbs on either side (e.g.\ \cite{gm65}; \cite{wg68};
\cite{rc76}; \cite{kc87}, hereafter KC87; \cite{rmk+88}; \cite{cks+92}).
It is generally accepted that bilateral SNRs represent an underlying
cylindrical symmetry, the emitting regions corresponding to the curved
walls of the cylinder (KC87). A bilateral appearance is then produced
when the ``barrel'' axis is approximately in the plane of the sky, the
two bright flanks running parallel to this axis.

Many mechanisms for generating a bilateral appearance have been suggested
(see \cite{bs95b} for a review). However it is not clear which of these
models, if any, actually apply.  Indeed, given the diversity in the
environments and ages of Galactic SNRs, several of the mechanisms proposed
may be valid.  We emphasise that observations demonstrate
that many SNRs are not bilateral. To some extent, this can be
accounted for by viewing angle (see KC87 and Section~\ref{nonbilateral}
below), but it is important to keep in mind that whatever mechanism
is invoked, the morphology may be dominated by small scale effects such as
the inhomogeneity of the local environment.

We now briefly outline previously proposed explanations for bilateral
SNRs. Explanations can be divided up into two
categories: ``extrinsic'', relating to the ambient ISM and magnetic field,
and ``intrinsic'', relating to the progenitor, its CSM, the SN explosion 
itself, and any compact stellar remnant.

\subsection{Extrinsic explanations}
\label{intro_extrinsic}

The simplest model to explain a bilateral appearance is that it is the
result of the density structure of the surrounding ISM.
Simulations show that when
a remnant expands inside an elongated, or tube-like, cavity,
it quickly takes the shape of that cavity, and a bilateral
appearance can be produced (\cite{bls90b}).

It has long been realised that the geometry of the ambient 
magnetic field can also affect the appearance of a SNR.
We collectively refer to models relying on the ambient field as
``magnetic''.
In a uniform magnetic field, the shock front produced by
an expanding SNR will
preferentially compresses the field where the shock
normal is perpendicular to the field lines. Thus, in a well-ordered
ambient field with a significant component in the plane of the
sky, enhanced emission along two limbs is generated, producing a
bilateral appearance (\cite{laa62b}; \cite{wg68}).

Argument is divided as to whether this mechanism can produce the
observed ratio of maximum to minimum intensity around the circumference
of SNRs which are in the adiabatic phase of evolution 
(\cite{fr90}, hereafter FR90; \cite{ram94}).
Thus, an extension of this model has been proposed, in which 
the efficiency of particle
acceleration by the SN shock depends on the angle between the shock
normal and the orientation of the ambient magnetic field 
(\cite{jok87}; \cite{rmk+88}; FR90\nocite{fr90}). FR90 argue that the bilateral
appearance in the adiabatic phase can best be explained by
``quasi-perpendicular'' acceleration, in which acceleration is most
efficient (and synchrotron emission strongest) where the
shock normal and the field are at right angles. ``Quasi-parallel''
models (where the shock normal and the field are aligned) can also
produce a bilateral appearance at some orientations, but at other
orientations produce morphologies which are never observed (FR90, but
see \cite{zwc96}).

\subsection{Intrinsic explanations}
\label{intro_intrinsic}

There are a number of symmetry axes associated with the progenitor
system and the explosion. Several mechanisms have been suggested
by which these symmetries can produce a bilateral appearance. Models
include a toroidal distribution of ejecta (\cite{bw83}; KC87\nocite{kc87}), 
the effect of a high velocity progenitor
(\cite{rtfb93}; \cite{bd94}), the distribution of mass
loss and magnetic field in the CSM produced by the progenitor
(\cite{man87}; \cite{its92}; \cite{ssmk92}; \cite{zwc96}),
and the influence of outflows from a central compact object
(\cite{md83}; \cite{wwps96}).

\subsection{A statistical approach}
\label{intro_stats}

Understanding the morphology of an individual SNR can be
difficult, often requiring a concerted multi-wavelength campaign.
Only a few bilateral SNRs can be considered to be well-studied
(most notably G296.5+10.0 and G327.6+14.6) and even these
objects provoke controversy
(e.g.\ compare Roger \etal 1988 with Storey \etal 1992\nocite{ssmk92}).
An alternative approach is to consider the common properties of
bilateral SNRs on a statistical basis. 

There is reasonable evidence that at least for low latitudes,
the Galactic magnetic field is well-ordered and generally runs parallel to
the Galactic Plane when projected onto the sky (\cite{mf70}; \cite{man74};
\cite{ea78}; \cite{sf83}; \cite{rk89}; \cite{rs90}; 
see also reviews by \cite{sfw86} and \cite{bbm+96b}). 
If a bilateral appearance is caused by a magnetic model, one might
expect the structure of the Galactic magnetic field to manifest itself
as a statistical alignment of the symmetry axes of bilateral SNRs with the 
Galactic Plane (\cite{sha69}).

Given that such a result would be reasonably conclusive, several
researchers have attempted to determine whether such an alignment
exists.  By creating an average circumferential profile from SNRs for
which appropriate data was available at the time, Shaver
(1969)\nocite{sha69} found that SNRs appeared to have minima in their
emission where the shell is intersected by a line of constant Galactic
latitude, and argued that this was evidence for magnetic models.
However, the statistical significance of the result was low, 
and the limited sample available was from observations of poor
resolution and sensitivity.

In more recent years, the statistics of
SNR samples have steadily improved, and other researchers have carried out
various analyses investigating alignment between the axes of
SNRs (defined in various ways) and the Galactic Plane 
(\cite{sha82}; \cite{man87}; \cite{lsc89}; \cite{wg96}, hereafter WG96).
In all cases, no alignment was found, leaving the authors to conclude
that either the magnetic field is not well-aligned with the Plane at such
scales, or that the magnetic model is not valid.

However, some of these studies implicitly assume that {\em all}\ SNRs
have a bilateral symmetry, and then assign an axis to every remnant in
their sample. Others restrict their sample to SNRs with a symmetry
axis, but include remnants where this symmetry is not necessarily
bilateral.  If only a small subset of remnants are truly bilateral,
then these analyses will never show any alignment because any
real effect will be swamped by the large number of randomly oriented axes
corresponding to non-bilateral SNRs.  In addition, the low
resolution of many of the images available at the time caused a poor
determination of the axis of symmetry.

There are now over 200 SNRs known in the Galaxy (\cite{gre96b}), and an
increasing number of objects have been observed at high resolution.
For almost the entire Galactic population, observations are now of
sufficient quality that one can determine which SNRs have a clear
bilateral appearance (defined in some manner), and restrict analysis to
this sample.

In Section~\ref{sec_observations}, we describe high-resolution 1.4 GHz
observations and analysis procedures for two Southern SNRs, \snra\
(formerly G003.7--00.2) and \snrb\ (formerly G350.0--01.8). We present
images of these remnants in Section~\ref{sec_results}, and argue that
these SNRs are members of the bilateral class with their axes aligned
along the Galactic Plane. The properties of these two SNRs are then
briefly discussed in Section~\ref{sec_discussion}.  In
Section~\ref{sec_bilaterals}, we consider the Galactic population of
bilateral SNRs, and demonstrate a clear tendency for the bilateral axis
to align with the Plane.  It is argued in Section~\ref{models}
that the magnetic model can
explain this alignment, but that there are difficulties accounting for
the detailed morphologies of bilateral SNRs.  An alternative model is
proposed, in which the ISM is pre-processed by the ambient magnetic
field to form structures running parallel to the Galactic Plane. SNRs
then interact with this material, producing a bilateral appearance with
the observed alignment.

\section{Observations and data reduction}
\label{sec_observations}

Interferometric observations were made with the Very Large Array (VLA)
of the National Radio Astronomy Observatory\footnote{The National Radio
Astronomy Observatory is a facility of the National Science Foundation
operated under cooperative agreement by Associated Universities, Inc.},
an aperture synthesis telescope consisting of 27 antennae in a
``Y''-shaped array, located near Socorro, New Mexico (\cite{nte83}).
The dates and lengths of the observations are given in
Table~\ref{tab_vla}.  Two bands were observed simultaneously, with
centre frequencies of 1.385~GHz and 1.465~GHz, bandwidths of 50 MHz, in
all four Stokes parameters. Because of its large angular extent, SNR
\snrb\ was observed in a mosaic of four pointings (see
Table~\ref{tab_snr}), with a cycle time of 24 minutes.  Amplitudes were
calibrated by assuming flux densities at 1.385 and 1.465~GHz
respectively of 14.94 and 14.55 Jy for J1331+305 (3C286) and 16.32 and
15.62 Jy for J0137+331 (3C48), where 1~Jy $= 10^{-26}$ W m$^{-2}$
Hz$^{-1}$.  Phases were calibrated using the source J1751--253.

The VLA data were edited and then calibrated in {\sc AIPS}
using standard techniques (\cite{gre96c}). The visibility data were then 
transferred to the Miriad package (\cite{stw95}), and
total intensity images formed using uniform weighting and a cell
size of 4$''$. For the mosaiced observations of \snrb, a joint
image was formed from all four pointings (\cite{chu93}; \cite{ssb96}), 
which was then corrected for the mean primary beam response of the 
VLA antennae.

The smallest projected spacing between antennae for the observations in
Table~\ref{tab_vla} is 0.12~k$\lambda$ , which corresponds to an
angular scale of $\sim 30'$. Thus the VLA observations are not sensitive
to emission on scales larger than this. These short spacings were
obtained by making single-dish observations (in total intensity only)
of both SNRs using the Parkes 64m radio telescope\footnote{The Parkes
radio telescope forms part of the Australia Telescope, which is funded
by the Commonwealth of Australia for operation as a National Facility
managed by CSIRO.}. 
Amplitudes were calibrated using the source PKS~B1934--638, assuming a
flux density of 14.90 Jy.  Except where noted in Table~\ref{tab_pks},
observational techniques were the same as used in the 2.4 GHz survey of
Duncan \etal (1995)\nocite{dshj95}.  Scans were made in both $l$ and
$b$, of regions surrounding both SNRs, at a scan rate of 6$^{\circ}$
min$^{-1}$.  The Parkes observations were reduced in the manner
described by Duncan \etal (1995)\nocite{dshj95}, and the resultant
images regridded and reprojected to match the higher resolution data.

Data from the two telescopes were then combined by adding together the
dirty images from each instrument. The flux density scale of the single
dish data was preserved by weighting them by the inverse ratio of the
beam areas (\cite{ytk91}; \cite{schn93}).  Point spread functions were
produced similarly, and the images then deconvolved using the maximum
entropy algorithm (\cite{gd78}). For \snrb, all pointings were
deconvolved simultaneously using the Miriad task MOSMEM (\cite{ssb96}).
The resulting models were then smoothed using a Gaussian restoring beam
to produce the final images.

Because of the proximity of \snra\ to both the Galactic Plane and
the Galactic Centre, single-dish observations of this source are
dominated by diffuse emission from confusing sources and the Galactic
background.  The composite image has flux filling the entire
field of view, and it is thus not possible to successfully deconvolve
it. The image of G003.7--00.2 as presented in Section~\ref{sec_results}
therefore does not include the Parkes measurements.  Since the angular
extent of this SNR is significantly smaller than the largest spatial
scale sampled by the VLA, we still expect to recover almost all of the
flux density from this object.

Images of linearly  polarized emission were made from the interferometric
data in Stokes Q, U and V.  (Although no real emission is expected in
circular polarization, the V image provides a useful check of
quality.)  The Q and U images were then deconvolved
using the CLEAN algorithm (\cite{cla80}), and a linear polarization
image, L, produced which was then corrected for 
non-Gaussian noise statistics (\cite{kbe86}) using the Miriad task IMPOL.

\section{Results}
\label{sec_results}

Total intensity images of \snra\ and \snrb\ are shown in
Figures~\ref{fig_g003} and~\ref{fig_g350} respectively.  
These images are of higher resolution,
and in the latter case, of greater extent than in previous images.
We are thus able to derive improved positions for the centres of these two
SNRs, resulting in a change of name for both remnants.

Because the VLA is non-coplanar, the approximation
of a flat sky used in standard synthesis imaging is invalid at
large distances from the phase centre (\cite{per89}). As a result, 
some point sources in Figure~\ref{fig_g350} 
have a triangular appearance.

Information about each image and its derived parameters are provided
in Table~\ref{tab_snr}.  Flux densities and their uncertainties were
determined by integrating within multiple polygons enclosing the
emission from each SNR.  The RMS noise in each image was measured by
similarly integrating over a source-free region. In the case of \snra,
the given centre and diameter are those of a circle best fitting the
outer perimeter of the emission.  \snrb\ is clearly non-circular, and
so the centre is defined as that point equidistant from the boundaries
of the SNR both along and at right angles to the bilateral axis. The
diameter given is the maximum extent in these two directions.

\subsection{\snra}

SNR \snra\ (then G003.7--00.2) was discovered in the Galactic Centre
Survey carried out with the Molonglo Observatory Synthesis Telescope
(MOST) at 0.843~GHz (\cite{gra94b}) and was classified as a SNR on
the basis of its morphology.  Using flux densities from Gray (1994) and
from Table~\ref{tab_snr}, 
we obtain a spectral index for \snra\ of $\alpha = -0.65
\pm 0.05$ ($S_{\nu} \propto \nu^{\alpha}$), confirming
the non-thermal nature of this source.

The image of \snra\ (Figure~\ref{fig_g003}) shows it to be a classic
bilateral remnant, with a clear axis of symmetry defined by
the two bright limbs, and a low level of emission along this axis
(cf.\ KC87).  The emitting regions appear to be composed of multiple
filaments extended along the symmetry axis:  on the north-west rim
of the SNR, these filaments take the form of roughly parallel arcs,
while in the south-east they overlap each other.  The original
0.843 GHz image of this remnant seemed to contain multiple overlapping
rings (\cite{gra94b}). However this appearance is most probably	due to the
alignment of various filaments when seen with the lower resolution of
the MOST --- no such features are evident in our image. 

$\psi$ is defined to be the acute angle between the symmetry axis of
the SNR and the Galactic Plane. We fit the symmetry axis by eye, and
estimate the uncertainty by fitting a set of possible axes to the
remnant. This gives a value $\psi = 6^{\circ} \pm 2^{\circ}$,
indicating close alignment between the symmetry axis and the Plane.
More involved methods for determining $\psi$ (e.g.\ \cite{ssmk92}) were
also implemented, giving similar results. For the accuracy required for
$\psi$ in this work, the more rigorous determinations are not
warranted.  Furthermore, for several sources considered in
Section~\ref{sec_bilaterals_subset}, images were not available in
electronic form.

Weak linearly polarized emission fills the entire image,
and it is not possible to determine how much
emission is specifically associated with the remnant. We put an
upper limit of 0.05~Jy on polarized emission from \snra, corresponding
to a fractional polarization of $<$3\%.

\subsection{\snrb}

SNR \snrb\ (then G350.0--01.8) was first identified from a 408 MHz survey of
the Galactic Plane carried out with the Molonglo Cross (\cite{gre72}).
At the comparatively low resolution and sensitivity of the time, only
the bright north-western arc was clearly delineated.  For the next 20
years, this remnant received little attention: Clark \etal
(1975)\nocite{ccg75} suggested it was ``part of an old large SNR'',
Caswell (1977)\nocite{cas77} noted it as an example of a SNR which was
brighter towards the Galactic Plane, while KC87 described it as
``difficult to classify''.  More recent observations at 0.843 GHz
(MOST; \cite{bb94}) and at 2.4
GHz (Parkes 64m; \cite{dshj95}) have demonstrated that this SNR is of
far greater extent than previously realised. Higher
resolution VLA observations are shown in Figure~\ref{fig_g350}.

Using a 2.4~GHz flux density of $18 \pm 1 $~Jy 
(obtained from the single-dish measurements of \cite{dshj95}),
and the 1.4~GHz value given in Table~\ref{tab_snr}, we derive a spectral
index for \snrb\ of  $\alpha = -0.4 \pm 0.1$. Note that the
flux density values quoted for this SNR by Clark \etal (1975)
\nocite{ccg75} correspond only to the bright north-west component.

The image of this SNR in Figure~\ref{fig_g350} shows a far more
complicated morphology than previously thought.  
It may even correspond to multiple remnants, a possibility discussed below
in Section~\ref{sec_discussion_g350}.
The brightest region of \snrb\ (corresponding to the previously
identified source G350.0--01.8, and labelled as A in Figure~\ref{fig_g350})
consists of several overlapping filaments, with the suggestion of
complex, unresolved, structure. This region, which extends over $\sim
40'$, forms a roughly circular arc with radius of curvature $\sim 55'$.
In the remnant's interior is region B, separated
from region A by a zone of diffuse emission of width $\sim 8'$.
Region B consists of a series of wisps, considerably fainter than
the emission in region A, but with a similar orientation and curvature.
A third region, C, is separated from region B by another zone of
diffuse emission of width $\sim 6'$. Region C
consists of still more overlapping filaments, of opposite
curvature to the rest of the SNR, which complete a distorted shell of 
emission. The outer rim of region C forms
a circular arc of radius $\sim 14'$, considerably different from the
radius of curvature for region A.

Although \snrb\ is clearly very different to \snra, it also
has a bilateral appearance. The circumference of the remnant is
bounded on two sides by regions of significant emission (namely A and C), and
at perpendicular position angles the emission is 
considerably weaker.   As for \snra, a bilateral axis can be defined.
This axis is shown in Figure~\ref{fig_g350}, 
drawn perpendicular to a line joining the
centres of curvature of regions A and C. For this axis,
$\psi = 5^{\circ} \pm 2^{\circ}$, again a high degree of
alignment with the Plane.

Clumps of linear polarization are detected throughout the interior
of \snrb. The total polarized intensity is 0.8 Jy, implying a fractional
intensity of 3.6\%.

\section{Discussion}
\label{sec_discussion}

Because of the lack of information at other wavelengths and
the proximity of both
remnants to $l = 0^{\circ}$,  the methods typically used to estimate
distances to SNRs cannot be applied (see \cite{gre84}).
The $\Sigma - D$ relation (e.g.\ \cite{cl79}; \cite{ht85}) has in the
past been used to estimate distances to SNRs,  but the uncertainties
are large (\cite{gre84}; \cite{ber86}). Thus we conclude that the age
of and distance to these SNRs remains unknown at the present time.

\subsection{\snra}

FR90\nocite{fr90} have quantified the contrast between the brightest
and faintest regions of a bilateral SNR by using an ``azimuthal
intensity ratio'', $A$. This is defined as the ratio between the peak
of emission around the SNR shell, and the minimum value around the
shell at the radius corresponding to this peak.  The filamentary nature
of this SNR makes it difficult to compute $A$ in this manner, but we
have made a commensurate calculation as follows. The shell of emission
from the SNR lies within an annulus of inner radius $4\farcm2$ and outer
radius $6\farcm7$, centred on the coordinates for \snra\ as listed in
Table~\ref{tab_snr}.  We have divided this annulus into 360 azimuthal
bins, and computed the average flux density in each bin. A peak of
approximately 1.1 mJy beam$^{-1}$ occurs at position angles (measured
north through east) of 100$^{\circ}$ and 260$^{\circ}$. The minimum
level of emission occurs at PA $\approx 40^{\circ}$, where no emission
can be detected above the 3$\sigma$ noise level. This corresponds to an
intensity ratio of $A > 6$, at a resolution of $\sim$70 beams per
diameter.  For a remnant in the adiabatic phase, this is consistent
with magnetic models for a bilateral appearance, but does not
distinguish between field compression alone and quasi-perpendicular
acceleration (FR90).

\subsection{\snrb}
\label{sec_discussion_g350}

One interpretation for the three spatially and morphologically distinct
regions of emission in \snrb\ is that they are two or three separate
SNRs, as has been postulated in objects such as G053.6-02.2 (3C400.2;
\cite{dggw94}) and 0547--697 (DEM L 316; \cite{wcd+97}).  Certainly
regions B and C are of similar surface brightness, which could
be interpreted as a complete shell, separate from region A.

If we consider a region on the sky defined by $330^{\circ} = -30^{\circ} \le 
l \le 30^{\circ}$, $1.5^{\circ} \le |b| \le 3^{\circ}$ (i.e.\ longitudes and
latitudes representative of \snrb), we find nine SNRs  in
the catalogue of Green (1996)\nocite{gre96b}.  Given this sky density, 
the probability
of finding two randomly distributed remnants less than $30'$ apart (as
required here) is 14\%. However, this is probably a lower limit,
since the small number of SNRs detected in this region
may be due to the incomplete sky coverage of
existing surveys. Also, SNRs are not necessarily randomly distributed,
but at least for those formed from massive stars probably represent
the clustering of their progenitors.  Thus there is no convincing
statistical argument that \snrb\ is a single SNR.

Nevertheless, the centres of curvature for all three regions lie in a
straight line, which bisects the three arcs of emission.  The fact that
regions A and C are of opposite concavity also suggests that all the
emission represents common expansion from a single event. Diffuse
emission from the SNR (contributed primarily by the single dish
observations) delineates a single plateau of emission bounded by
regions A and C.  We therefore consider the most likely interpretation
to be that the emission in this region is from a single object.

It is possible that the non-circular appearance could be due to the
influence of a pulsar or other compact object.  An analogy could be
drawn with SNR G005.4--01.2, which has a bright, flattened ridge of
emission along one side (\cite{ckk+87}; \cite{fkw94}), resembling
region A of \snrb.  It has been proposed that the distorted morphology
of this SNR is a result of the associated pulsar PSR~B1757--24
overtaking the expanding parent shell (\cite{fk91}). While a
flat-spectrum region extending from the pulsar back towards the radio
shell supports such a model for SNR G005.4--1.2 (\cite{fk91};
\cite{fkw94}), no similar component is observed in \snrb, nor has any
pulsar been detected (\cite{mdt85}; \cite{kmj+96}).  Thus we consider
an interaction with a compact object an unlikely explanation for the
observed morphology.  Given that any associated pulsar should be located
near the SNR's presumed centre of expansion (\cite{gj95b}), 
the area bounded by regions B and C might be a more fruitful target for 
future pulsar searches.

A 1.4 GHz image of SNR G166.0+04.3 (VRO 42.05.01; \cite{lprv82} ;
\cite{plr87}) is shown in Figure~\ref{fig_vro}, revealing a striking
resemblance between this SNR and \snrb.  G166.0+04.3 is also composed
of three separate regions, whose relative shape, size and orientation
are similar to those in \snrb. Both SNRs are bounded on one side by a
bright arc with a large radius of curvature which trails into the noise
(region A), contain a fainter strip of emission parallel to A and
separated from it (region B), and a series of filaments with a
distinctly smaller radius of curvature and of opposite concavity to the
rest of the emission, completing the outline of a shell (region C).
Both SNRs are similarly aligned to the Galactic Plane:  when an axis of
symmetry is defined for
G166.0+4.3, one finds that $\psi = 2^{\circ} \pm 2^{\circ}$.

There are notable differences between the SNRs,
however. Firstly, in \snrb, regions B and C are of similar brightness but are
much fainter than A, while in G166.0+04.3, the region corresponding
to C (the ``shell'' of \cite{lprv82}) is of comparable brightness
to that corresponding to region A (the ``wing''). Secondly,
the central region of G166.0+04.3 consists of linear features, while
region B appears to be curved. Finally, while in
\snrb\ regions B and C are separated by a region of faint emission, the
analogues of these two regions in G166.0+04.3 actually overlap.

Pineault \etal (1987)\nocite{plr87} explain the morphology of G166.0+04.3 
in terms of 
the interaction of an expanding remnant with a slab of pre-existing low
density material. The SN explosion is believed to have occurred at or
near the centre of curvature of the ``shell'' component, which
represents the expansion of that part of the SN shock into a uniform
medium.  The faint linear features in the remnant's centre mark the
re-energising of the slab's surface as a result of the shock breaking
out into it, and the ``wing'' component represents the shock once
again encountering dense material on the other side.

The similarity between the two remnants suggests a similar explanation
for them as well. In this model, region C of \snrb\ 
is the unperturbed component of the 
original SNR, which evolves
independently of those parts of the remnant which encounter
a more complicated structure (\cite{fg82}; \cite{tby85}). 
Indeed, region C resembles
SNR G119.5+10.2 (CTA 1; \cite{plmg93}), a partial shell for which
there is evidence that a breakout has occurred but for which the
remainder of the SN shock is not visible. Region B of \snrb\ represents the
boundary of a low density slab. That it is concave and
not linear and does not overlap with region C, suggests that the details
of the 3D geometry differ from that in the case of G166.0+04.3. Region A
is that component of the shock which has traversed the slab and continues
to propagate beyond it.

\section{The Galactic population of bilateral SNRs}
\label{sec_bilaterals}

\subsection{Selection criteria}

Although previous studies have argued that no alignment exists between
the axes of bilateral SNRs and the Galactic Plane, it is interesting
that \snra, \snrb\ and G166.0+04.3 should all have such low values of
$\psi$. This similarity prompted an examination of other bilateral
SNRs.

We have therefore reviewed the Galactic SNR catalogue of Green (1996),
which currently contains 215 remnants.  The criteria for a SNR to be
classed as bilateral are as follows:

\begin{itemize}

\item{} It must be of the shell or composite class.
\item{} The highest resolution image available must have a minimum of 
10 beams across its diameter.
\item{} The SNR must have clear minima in emission separated by 
position angles of $180^{\circ}\pm 30^{\circ}$ relative to the assumed 
centre of the SNR. The presence of opposed minima is the main
criterion for classification, as it is these minima which
define the bilateral axis.
\item{} The SNR must have well-defined maxima, similarly opposed, and
at approximately perpendicular position angles to the minima.
\item{} A clear bilateral axis should be identifiable, passing through
the two minima and through the centre of the SNR.  Objects with a large
uncertainty ($\gtrsim20^{\circ}$) in the position angle of the axis are
excluded --- this requirement is critical in defining the objects in
the sample.

\end{itemize}

We do not quantify the azimuthal
intensity ratio, $A$, of FR90 as a threshold.
Such a measurement is
both difficult and misleading, because of the various presentations of
images in the literature, and of the differing nature of the
observations involved.  As FR90 have pointed out, the value of $A$
depends strongly on the resolution of an image, so that the appearance of
bilaterality will increase with resolution.  Also, the lack of
large-scale emission present in some interferometric images can cause a
further over-estimation of this parameter.  Thus using this ratio
as a threshold will not result in a consistent criterion.

\subsection{Non-bilateral SNRs}
\label{nonbilateral_obs}

There are 198 SNRs in Green's (1996)\nocite{gre96b} SNR catalogue which
do not meet our criteria.  This is in strong contrast to KC87, who
found that 63\% of all SNRs had some element of ``barrel'' structure.
This primarily reflects the strictness of our criteria.
For example, consider G315.4--02.3 (RCW~86; Figure~\ref{fig_g315})
and G332.4--00.4 (RCW~103; Figure~\ref{fig_g332.4}), both of which
are described by 
KC87 as ``well developed barrels''. The former has no opposed
maxima or minima. The latter is a marginal case (Dickel \etal 1996
\nocite{dgym96} describe it is an ``incipient barrel''), but lacks
the distinctive minima and symmetry properties 
of the best examples of the class.

Improved observations over the last ten years have allowed a more precise
classification of many remnants. For the MOST, substantial
improvements to both hardware (Amy \& Large 1990\nocite{al90}, 
1992\nocite{al92}) and data reduction
techniques (\cite{cy95}) since the observations of KC87
were made have significantly reduced the effect of artifacts, increased
the sensitivity, and generally improved image quality.

\subsection{Bilateral SNRs}
\label{sec_bilaterals_subset}

There remain 17 remnants in the catalogue which can be classed as
clearly bilateral from our criteria.  A list of these bilateral SNRs is
given in Table~\ref{tab_psi}.  For each remnant a bilateral axis is
defined by eye, as for the SNRs in Section~\ref{sec_results}.  This
axis passes through the two minima, corresponding closely to an axis of
mirror symmetry.  Some examples of SNRs which we have classed as
bilateral are shown in Figures~\ref{fig_g46} through~\ref{fig_g356},
with the bilateral axis of each remnant shown. In each case, we have
measured the acute angle, $\psi$, between this axis and the Galactic
Plane.  Values of $\psi$ for each SNR are shown in
Table~\ref{tab_psi}.

The reader is encouraged to examine these (and other) SNRs in order to
verify that the classifications made are reasonable.  We should
emphasise that if there is any doubt, a SNR is {\em excluded} from our
class. If a preferential alignment exists, it will remain apparent if
some bilateral SNRs are excluded from the sample (provided that the
statistics are still meaningful), but will be masked if too many
remnants which do not satisfy the criteria are included.

\subsection{Orientation with respect to the Galactic Plane}

A histogram of the distribution in $\psi$ from Table~\ref{tab_psi} is shown in
Figure~\ref{fig_psi}.  
This distribution has a median of 
$\psi_m = 12^{\circ}{}^{+3^{\circ}}_{-2^{\circ}}$,
where the uncertainty corresponds to the maximum shift in median
should two objects be discarded from the sample. This result
suggests an alignment of bilateral SNRs with the Galactic Plane:
the probability of this distribution occurring by chance is now discussed.

Assume cylindrical symmetry for bilateral SNRs, and represent the
three-dimensional axis of such a remnant by a vector in Cartesian
space, with spherical coordinates $\theta$, $\phi$ 
as shown in Figure~\ref{fig_axes}.  
The observer views the system along the $r$ axis, and the $l$ and $b$
axes represent increasing Galactic longitude and latitude respectively
in the plane of the sky.  Note that the observed axis is a projection
onto the $l-b$ plane of the true axis, and that a remnant will only
have a bilateral appearance provided that $\theta > \theta_c$, where
$\theta_c$ depends on the details of the geometry (KC87).
We only consider values in the range $|\phi| \le 90^{\circ}$, since the
other hemisphere is equivalent.  Randomly oriented vectors will have a
uniform distribution in $\phi$ and hence in $\psi$. Thus all values of
$\psi$ are equally likely, independent of $\theta_c$.

Using the binomial theorem, we can calculate that the probability of a
randomly oriented sample having the observed median is
$0.7^{+4.7}_{-0.1} \times 10^{-3}$. We argue in
Section~\ref{sec_bilaterals_highlat} that G296.5+10.0 and G327.6+14.6
may be distinct from the other objects in Table~\ref{tab_psi}, as a
consequence of their large distance from the Plane. These two remnants
have the highest values of $\psi$, and if they are excluded from the
sample, the probability is even more tightly constrained.

This calculation shows that the probability of the distribution being
randomly oriented is low. {\em We therefore consider our data to
demonstrate strong evidence that bilateral SNRs align with the Galactic
Plane.}

Given that no correlation between the Galactic Plane and the rotational 
axes of SNR progenitors has been observed (\cite{sle49}; \cite{hs54}), 
intrinsic models for the bilateral appearance (as listed in
Section~\ref{intro_intrinsic}) are not consistent with  the observed
distribution of $\psi$.  These models
are probably better suited to explain the
appearance of other classes of remnant morphology.  For example,
a toroidal explosion may be appropriate for G292.0+01.8 (\cite{tcb82}),
a high velocity progenitor could describe G357.7-00.1 (\cite{ssp+85};
\cite{bh85}), outflows from a central source are appropriate for
G039.7-02.0 (W50 and SS433; \cite{eb87}), while the properties of the
CSM perhaps only cause a bilateral appearance at an early age, such as
is observed in the radio remnant of SN~1987A (\cite{gms+97}).

\section{Models for bilateral SNRs}
\label{models}

\subsection{A magnetic model}
\label{sec_bilaterals_mag}

As discussed in Section~\ref{intro_stats}, an alignment of bilateral
SNRs can be interpreted as evidence for magnetic
models (\cite{sha69}; \cite{man87}).  In order for such mechanisms
to be effective, the surrounding magnetic field must be
well-ordered and the ISM into which the SNR is expanding must be
reasonably homogeneous. The small number of bilateral remnants may indicate
that this condition rarely occurs, as will be discussed further below.

There is some support for the magnetic model from polarization
measurements. Specifically, in the field compression scenario
(\cite{wg68}), one expects the magnetic field direction in the emitting
regions to be oriented tangentially. Such a phenomenon is observed in
the bilateral SNRs G296.5+10.0 (\cite{mh94b}) and G093.3+06.9
(\cite{lsmt84}). A tangential magnetic field need not always be observed,
however, since in younger SNRs other factors may cause the magnetic
field to be oriented radially (\cite{rg93}; \cite{jn96}) or be
generally disordered (FR90; \cite{mr94}).

One argument against magnetic models is that the typical energy density
in the ambient magnetic field is many orders of magnitude less than the
kinetic energy of a typical SNR shock (\cite{man87}; \cite{bls90b}).
However these arguments only pertain to whether the {\em shape} of a
remnant can be influenced by the surrounding field. The brightness
distribution can be significantly affected by the ambient field, even
if the field is dynamically unimportant (FR90).

\subsection{An alternative to the magnetic model}

While magnetic models can simply explain bilateral SNRs which have a
circular rim and two opposed limbs of similar shape and brightness
(e.g.\ \snra\ and G327.6+14.6), we now argue that difficulties are
encountered when one considers whether such models can also
account for the detailed morphologies of other objects in
Table~\ref{tab_psi}. As will be discussed in 
Section~\ref{sec_bilaterals_ism}, an alternative explanation is
the effect of the ISM.

\subsubsection{Asymmetric SNRs}

There are objects in Table~\ref{tab_psi} which align closely
with the Plane, but for which the flux density and morphology of the
bright opposed limbs is very different. The most striking examples of
this class are G320.4--01.2 (Figure~\ref{fig_g320}) and G308.8--00.1,
but this asymmetry is also present to a lesser extent in G078.2+02.1,
G127.1+00.5, \snrb\ (Figure~\ref{fig_g350}) and G356.3--01.5
(Figure~\ref{fig_g356}).  In the magnetic model, this morphology can be
explained by a gradient in the field strength across the remnant
(\cite{cl79}; \cite{rf90}).

Alternatively, for both G320.4--01.2 and G308.8--00.1, it has been
suggested that this appearance is due to the influence of a central
pulsar on the surrounding shell (\cite{md83}; \cite{kmj+92};
\cite{bb97}), or because one side of the remnant encounters a region of
higher density (\cite{shmc83}; \cite{kmj+92}).  In the latter case, the
low value of $\psi$ for these remnants suggests the presence of density
gradients perpendicular to the Plane.  There is no clear tendency for
the limb closest to the Plane to be the brighter of the two. This is
contrary to the suggestion of Caswell (1977)\nocite{cas77}, although he
noted that this may only apply for remnants at large $|z|$.

\subsubsection{Distorted SNRs}

Some objects are significantly distorted from a circular or elliptical
shape, for example \snrb\ (Figure~\ref{fig_g350}) and G166.0+04.3
(Figure~\ref{fig_vro}).  Observations of the latter at radio (Landecker
\etal 1982, 1989\nocite{lprv82,lprv89}), X-ray (\cite{bg94};
\cite{gb97}) and optical (\cite{ppl+85}) wavelengths, are all
consistent with the model of Pineault \etal (1987)\nocite{plr87}. As
discussed in Section~\ref{sec_discussion_g350}, this explains the
morphology in terms of the shock encountering a hot low density tunnel
lying along the Galactic Plane, before propagating into dense material
beyond.  We have noted that \snrb\ has a similar appearance, which by
analogy might also be explained in terms of such a model.  The gross
morphology of these remnants can be reproduced by magnetic models, but
an abrupt change in density by a factor of 100, along an interface
parallel to the Galactic Plane, is still required (\cite{ms90}).

\subsubsection{Elongated SNRs}

Some objects in Table~\ref{tab_psi} are significantly extended along
their symmetry axis, most notably G356.3--01.5 (Figure~\ref{fig_g356})
and G296.5+10.0 (Figure~\ref{fig_g296}).  This appearance can be
explained by tension associated with the ambient magnetic field, which
would cause remnants to preferentially expand along field lines
(\cite{ir91}; \cite{mss93}; \cite{rt95}).  However for typical values
of the explosion energy and ambient field strength, 
the shape of a SNR will not be appreciably
distorted by the magnetic field until it reaches a diameter of
$\gtrsim$100~pc (\cite{ms90}; \cite{nor93}), at which stage a remnant is
nearing the end of its observable lifetime (e.g.\ \cite{sfs89}).  In
order for tension in the magnetic field to account for the observed
extension, one must invoke a low explosion energy ($\lesssim 10^{50}$~erg)
and/or a high magnetic field ($\gtrsim 100$~$\mu$G).  An alternative
magnetic explanation for this elongation has been proposed by Roger
\etal (1988)\nocite{rmk+88}.  In the case of G296.5+10.0, they
speculate that the SNR shock may not couple to newly swept up material
when the magnetic field is parallel to the direction of propagation.
As the remnant evolves the shock speed is dependent on the angle with
the magnetic field and the remnant will elongate.

Bisnovatyi-Kogan \etal (1990)\nocite{bls90b} consider a non-magnetic
explanation for G296.5+10.0, wherein both the elongation and the
bilateral appearance of the SNR are the result of it having expanded
into a similarly extended pre-existing cavity.  For this mechanism to
apply for SNRs with low values of $\psi$, however,
such cavities must lie with their major axis along the Plane.

\subsubsection{An ISM model}
\label{sec_bilaterals_ism}

In the above discussion, we have considered three classes of objects
which can be more comfortably accounted for by the surrounding ISM 
than by the magnetic model.  We have considered objects with one side
brighter than the other as might be due to a density gradient, a
breakout morphology due to the presence of hot tunnels, and elongated
remnants which might be due to an extended cavity.  In order for any
of these explanations to be valid, the observed distribution
of $\psi$ requires that structures in the ISM must be consistently
parallel to the Plane.

As a mechanism for the bilateral appearance, we propose that the Galactic 
magnetic field produces density stratifications parallel to the Galactic 
Plane.  Evidence to support this hypothesis is now presented.

The first issue to consider is whether stellar wind-bubbles can be
elongated by the surrounding field.  Consider a homogeneous ISM of gas
density $n_H$, in which there is a uniform magnetic field of strength
$B_0$. The bubble produced by an isotropic wind will expand more
rapidly along field lines than across them, and will begin to be
significantly distorted when the internal pressure in the bubble falls
below the external magnetic pressure (\cite{kon82}). Adapting the
result of Stone \& Norman (1992)\nocite{sn92} for a wind from a
proto-star, this will occur after a time $t_1$, given by :

\begin{equation}
\left( \frac{t_1}{Myr} \right) \approx  2
\left( \frac{B_0}{3\, {\rm \mu G}} \right)^{-5/2}
\left( \frac{L_w}{10^{21}\, {\rm erg}} \right)^{1/2} 
\left( \frac{n_H}{\rm cm^{-3}}
\right)^{3/4},
\label{eqn_wind}
\end{equation}

\noindent where $M$ is the mass of the star, $v_w$ is the velocity of
the wind and $L_w = \frac{1}{2} \dot{M} v_w^2$ is the associated
mechanical luminosity.
Therefore as a rough criterion, a wind bubble will become significantly
elongated along ambient field lines if $t_0 \gg t_1$, where $t_0$
is the main-sequence lifetime of the star.

Chevalier \& Liang (1989)\nocite{cl89} tabulate the properties of
stellar wind-bubbles for different spectral types, from which we find
that in a typical environment ($B_0 = 3$~$\mu$G, $n_H = 1$~cm$^{-3}$),
a main-sequence wind-bubble will be dominated by the ambient field for
a star of spectral type O9 or later. Such stars comprise the vast
majority of the stellar population, and indeed Lozinskaya
(1988)\nocite{loz88} notes that wind-blown bubbles are often
elongated.  Since $t_1 \propto B_0^{-5/2}$, only a slightly higher
field than assumed above ($\gtrsim 10$~$\mu$G) is required to distort
wind-bubbles produced by even the most massive stars.
Note that the extra energy input
in the late phases of stellar evolution may complicate the simple
picture presented here.

On larger scales, Heiles (1979)\nocite{hei79} has found that
\HI\ shells within $10\arcdeg$ of the Plane appear elongated in the
longitudinal direction, which he suggests could
also be due to the ambient field.
Also, simulations of magnetised superbubbles
(Tomisaka 1990, 1992\nocite{tom90}\nocite{tom92}; \cite{mss93}) 
and old SNRs (\cite{nor93}) show that
such structures also become elongated along field lines, but over larger
length and time scales than for stellar wind bubbles.

To summarise, we propose that where the magnetic field is oriented
parallel to the Galactic disc, structures will be produced extended in
the longitudinal direction, on approximately the same scale as SNRs.
Thus close to the Plane, the structure of the Galactic ISM might take the
form of tunnels, gradients and cavities all elongated in the disc of the
Galaxy. When a SNR interacts with this structure, a bilateral appearance
is produced (as in the model proposed by \cite{bls90b}), with the
axis aligned parallel to the Plane.  

A remnant formed from a massive progenitor will interact at its
earliest stages with a complex and extensive CSM.  During this period,
a SNR can take on some aspects of bilaterality (\cite{its92};
\cite{blc96}; \cite{zwc96}; \cite{gms+97}), but no alignment with the
Galactic Plane will be observed (unless the progenitor wind-bubble is
elongated along the Plane through the process discussed above).  Only
once the shock propagates beyond its CSM and begins to interact with
the ambient ISM should any trend in $\psi$ manifest itself. Type Ia
SNRs such as G327.6+14.6 may interact with the ambient medium at a much
earlier stage, however.  At the other end of the scale, as a remnant
ages and expands, it becomes more and more likely to encounter complex
inhomogeneities in the ISM which will distort its appearance.  Hence,
this mechanism for the bilateral appearance is most relevant for
middle-aged remnants, rather than young or old ones. The uncertainty in
SNR age estimates prevents us from demonstrating this property in the
observed sample. However rough estimates for the ages of SNRs in
Table~\ref{tab_psi} (using the $\Sigma-t$ relation; Caswell \&
Lerche 1979\nocite{cl79}) give that most are between 1\,000 and
10\,000 years old.  

\subsection{Exceptional cases: SNRs G296.5+10.0 and G327.6+14.6}
\label{sec_bilaterals_highlat}

SNRs G296.5+10.0 (Figure~\ref{fig_g296}) and G327.6+14.6
(Figure~\ref{fig_sn1006}) are discrepant in their values of $\psi$,
both being oriented at almost right angles to the Plane.  However, these two
remnants are also notable in that
they are both at a significant height above the Plane (175--350~pc and 
450~pc, respectively; \cite{rmk+88}).  

For bilateral SNRs at low latitudes, we have proposed ISM structures
running parallel to the Plane.  At higher latitudes, the ISM is
dominated by perpendicular structures in the form of chimneys
(\cite{ni89}; \cite{ntd96}) and ``worms'' (\cite{hei84};
\cite{khr92}).  Stellar polarization measurements demonstrate the
existence of large-scale loops of magnetic field also extending
perpendicular to the Plane (\cite{mf70}).  Thus that bilateral SNRs
might tend to have $\psi \sim 90\arcdeg$ at large distances from the
Plane is consistent with the extrinsic models described in Sections
\ref{sec_bilaterals_mag} and \ref{sec_bilaterals_ism}.  SNRs
G296.5+10.0 and G327.6+14.6 can therefore still be incorporated within
the proposed model.  One can speculate that bilateral SNRs at
intermediate scale heights might be in transition between remnants with
$\psi \sim 0^{\circ}$ and those for which $\psi \sim 90^{\circ}$.
However the uncertainty in the distances to SNRs in our sample prevents
us from making meaningful conclusions on any latitude dependence
at present.

These two SNRs are regarded as the prototypes for the bilateral
morphology (e.g.\ KC87; \cite{rmk+88}). That the most striking examples
of bilaterality constitute 50\% of known remnants for which $|b| \ge
10^{\circ}$ may be indicative of the homogeneity of the environment far
from the Plane. It is highly likely that the radio all-sky surveys
currently being carried out (\cite{lcc+94}; \cite{ccg+96}) will
discover new high latitude SNRs. Whether such remnants also have a
well-ordered appearance and a high degree of symmetry will be useful
(but not conclusive) evidence regarding the model we have proposed.

We note that intrinsic models have also been proposed for these two
SNRs:  Storey \etal (1992) argue that the symmetry present in
G296.5+10.0 over a wide range of spatial scales generally argues
against magnetic models for this object, while Willingale \etal (1996)
hypothesise that a central compact object with opposed jets could
generate the morphology of G327.6+14.6.  Whether these remnants
represent an extrinsic mechanism operating at high Galactic latitude,
or intrinsic effects, either way, their high value of $\psi$ is not
at odds with the alignment observed in the remainder of the sample.

\subsection{Explanations for non-bilateral structure}
\label{nonbilateral}

As noted in Section~\ref{nonbilateral_obs}, 
the vast majority of remnants do not have a clear bilateral
appearance.  The wide range of SNR morphologies observed (clearly
demonstrated in the catalogue of WG96) is 
indicative that complex, small-scale effects can dominate the
appearance of SNRs. Specifically, the absence of bilaterality can
be explained by any of the following conditions:

\begin{itemize}

\item{} A dominant component of the field along the line of sight, so
that the projection of the emitting region is no longer two distinct
arcs of emission. KC87 estimate that a SNR with a cylindrical 3D
geometry will appear as a uniform ring of emission with no bilateral
appearance if $\theta < 25^{\circ}$, and will be ``confused'' if
$25^{\circ} < \theta < 45^{\circ}$. This results in 24\% (rounded to
30\% in KC87) of SNRs having a non-bilateral appearance, simply due to
orientation.

\item{} An inhomogeneous ISM. In this case,  encounters with molecular
clouds, cavities, and other inhomogeneities distort a remnant
sufficiently so that no symmetrical structure can be observed
(e.g.\ \cite{plmg93}; \cite{rm93}; \cite{km97}).

\item{} A disordered ambient magnetic field, so that in the magnetic
model, no position angle around the circumference has enhanced
emissivity. It is suggestive that some bilaterals are highly polarized
(e.g.\ \cite{lsmt84}; \cite{mh94b}), indicating a well-ordered magnetic
field.

\end{itemize}

\section{Conclusion}

We have presented 1.4~GHz radio observations of two southern
supernova remnants, \snra\ and \snrb.  The
morphologies of these two SNRs are distinct --- \snra\ shows a
filamentary two-arc appearance in a classic ``barrel'' shape, while
\snrb\ strongly resembles the breakout morphology of G166.0+04.3 
(VRO~42.05.01).

Despite their differences, both remnants have a clear bilateral
appearance. Furthermore, the axis defined by this classification in both cases
aligns almost exactly with the Galactic Plane. On examination of the
entire sample of known Galactic SNRs, we find 17 SNRs with a high level
of bilateral morphology, of which nine are oriented to the Plane at 
$12^{\circ}$ or less. With a minimum of assumptions regarding the
three-dimensional geometry, the probability of this occurring by chance is
$7 \times 10^{-4}$. Considering SNRs G296.5+10.0 and G327.6+14.6 as
distinct from the remainder of the sample reduces this probability even
further.

This trend rules out ``intrinsic'' models for the bilateral morphology,
all of which involve an axis defined by the orientation or motion of
the progenitor star. Such models are not expected to produce remnants which
align in any consistent way with the Galactic Plane.

The general alignment of bilateral remnants with the Plane can be
explained by ``magnetic'' models, whereby an ambient magnetic field
generates bright regions around the SNR shell where the shock normal
is perpendicular to the field. This is a result of either enhanced
field compression or preferential electron acceleration in these
regions.  However, distorted
morphologies in bilateral SNRs can be better explained if the
ISM into which these remnants expand consists of structures
preferentially elongated in the longitudinal direction. We
propose that this occurs in the form of expanding wind-bubbles 
distorted by the surrounding magnetic field.

The majority of SNRs do not have a distinct bilateral appearance.
For some SNRs, this can be explained as an orientation effect, when
the bilateral axis is close to the line of sight. The effects of
a disordered magnetic field and an inhomogeneous ISM can further account
for the observed diversity in morphologies.

This study has demonstrated the value of high resolution observations of
Galactic SNRs. Further observations, in particular of the polarization
and environment of bilateral SNRs, are needed to test the proposed
models. Undoubtedly there also remain more examples of
bilateral SNRs to be discovered, particularly in the first quadrant and
at high Galactic latitudes, which will improve the available statistics. 

While theoretical modelling of the evolution of stellar wind bubbles
is quite sophisticated (\cite{glm96} and references therein),
the effect of external magnetic fields is generally ignored. The
simple calculations made here suggest that such fields can have a
significant effect on the dynamics of a bubble and should be
included in subsequent modelling.  Similarly, the interaction of SNRs
with surrounding material usually assumes spherical symmetry (\cite{cl89};
\cite{trfb91}); detailed simulations of the interaction of remnants with
the (non-spherical) structures proposed here would be highly desirable.

An interesting extension of the results described would be to
determine whether bilateral SNRs in other edge-on spiral galaxies are
aligned along preferred position angles, and whether these angles
correlate with measurements of the corresponding galactic magnetic
field. Of particular interest is NGC~4631 which largely has a magnetic
field oriented perpendicular to the disc (\cite{hbd91};
\cite{gh94}).  At present, existing observations of SNRs in other
spiral galaxies (e.g.\ \cite{dvgv93}; \cite{mpw+94}) are of
insufficient resolution to derive a bilateral subset.  However, the
planned upgrade to the VLA (\cite{bb96c}) will provide sufficient
improvement in sensitivity and resolution to
produce images of SNRs in these galaxies of a quality comparable to
existing Galactic observations.

\acknowledgements

I am grateful to Miller Goss and Dale Frail for their help in
preparing for the VLA observations, and for their assistance
with {\sc AIPS} and other aspects of data reduction.  I also
thank Roy Duncan for carrying out the observations made with
the Parkes radio telescope, and subsequently reducing and assisting
with these data. I would like to thank many people for interesting
discussions:  John Dickel, John Dickey, Anne Green, Jeff
Hester, Tom Landecker, Dick Manchester, Michael Norman, Steve Reynolds,
Rob Roger, Brad Wallace and Mark Wardle. 
David Moffett suggested the possibility that
stellar wind-bubbles might be influenced by the ambient magnetic field.
Gloria Dubner, Andrew Gray, Anne Green, Tom Landecker and Mike Kesteven
kindly provided images of various SNRs. I thank NRAO for hospitality
during my visit to Socorro, and acknowledge the financial support of
an Australian Postgraduate Award and of the Support for Access to Major
Research Facilities programme, administered by the Australian Department
of Industry, Science and Tourism.  This research has made use of the
NASA Astrophysics Data System and of the CDS SIMBAD database.

\clearpage 
\newpage

\bibliography{bilaterals}

\begin{thebibliography}{}

\bibitem[Amy \& Large 1990]{al90}
Amy, S.~W. \& Large, M.~I. 1990, { Proc. Astron. Soc. Austral.}, {\rm 8}, 308.

\bibitem[Amy \& Large 1992]{al92}
Amy, S.~W. \& Large, M.~I. 1992, { Aust. J. Phys.}, {\rm 45}, 105.

\bibitem[Bastian \& Bridle 1996]{bb96c}
Bastian, T.~S. \& Bridle, A.~H. 1996, The VLA Development Plan.

\bibitem[Beck \etal  1996]{bbm+96b}
Beck, R., Brandenburg, A., Moss, D., Shukurov, A., \& Sokoloff, D. 1996, {
  \araa}, {\rm 34}, 155.

\bibitem[Becker \& Helfand 1985]{bh85}
Becker, R.~H. \& Helfand, D.~J. 1985, { Nature}, {\rm 313}, 115.

\bibitem[Berkhuijsen 1986]{ber86}
Berkhuijsen, E.~M. 1986, { \aap}, {\rm 166}, 257.

\bibitem[Bisnovatyi-Kogan, Lozinskaya, \& Silich 1990]{bls90b}
Bisnovatyi-Kogan, G.~S., Lozinskaya, T.~A., \& Silich, S.~A. 1990, { \apss},
  {\rm 166}, 277.

\bibitem[Bisnovatyi-Kogan \& Silich 1995]{bs95b}
Bisnovatyi-Kogan, G.~S. \& Silich, S.~A. 1995, { Rev. Mod. Phys.}, {\rm 67},
  661.

\bibitem[Blondin, Lundqvist, \& Chevalier 1996]{blc96}
Blondin, J.~M., Lundqvist, P., \& Chevalier, R.~A. 1996, { \apj}, {\rm 472},
  257.

\bibitem[Bodenheimer \& Woosley 1983]{bw83}
Bodenheimer, P. \& Woosley, S.~E. 1983, { \apj}, {\rm 269}, 281.

\bibitem[Brazier \& Becker 1997]{bb97}
Brazier, K. T.~S. \& Becker, W. 1997, { \mnras}, {\rm 284}, 335.

\bibitem[Brighenti \& D'Ercole 1994]{bd94}
Brighenti, F. \& D'Ercole, A. 1994, { \mnras}, {\rm 270}, 65.

\bibitem[Burn \& Bush 1994]{bb94}
Burn, D. \& Bush, B. 1994, Private communication.

\bibitem[Burrows \& Guo 1994]{bg94}
Burrows, D.~N. \& Guo, Z. 1994, { \apjlett}, {\rm 421}, L19.

\bibitem[Caswell 1977]{cas77}
Caswell, J.~L. 1977, { Proc. Astron. Soc. Austral.}, {\rm 3}, 130.

\bibitem[Caswell \etal  1987]{ckk+87}
Caswell, J.~L., Kesteven, M.~J., Komesaroff, M.~M., Haynes, R.~F., Milne,
  D.~K., Stewart, R.~T., \& Wilson, S.~G. 1987, { \mnras}, {\rm 225}, 329.

\bibitem[Caswell \etal  1992]{cks+92}
Caswell, J.~L., Kesteven, M.~J., Stewart, R.~T., Milne, D.~K., \& Haynes, R.~H.
  1992, { \apjlett}, {\rm 399}, L151.

\bibitem[Caswell \& Lerche 1979]{cl79}
Caswell, J.~L. \& Lerche, I. 1979, { \mnras}, {\rm 187}, 201.

\bibitem[Chevalier \& Liang 1989]{cl89}
Chevalier, R.~A. \& Liang, E.~P. 1989, { \apj}, {\rm 344}, 332.

\bibitem[Clark 1980]{cla80}
Clark, B.~G. 1980, { \aap}, {\rm 89}, 377.

\bibitem[Clark, Caswell, \& Green 1975]{ccg75}
Clark, D.~H., Caswell, J.~L., \& Green, A.~J. 1975, { Aust. J. Phys. Astr.
  Supp.}, {\rm 37}, 1.

\bibitem[Condon \etal  1996]{ccg+96}
Condon, J.~J., Cotton, W.~D., Greisen, E.~W., Yin, Q.~F., Perley, R.~A.,
  Taylor, G.~B., \& Broderick, J.~J. 1996, available on the WWW at
  http://www.cv.nrao.edu/\~{ }jcondon/nvss.html.

\bibitem[Cornwell, Holdaway, \& Uson 1993]{chu93}
Cornwell, T.~J., Holdaway, M.~A., \& Uson, J.~M. 1993, { \aap}, {\rm 271}, 697.

\bibitem[Cram \& Ye 1995]{cy95}
Cram, L. \& Ye, T. 1995, { Aust. J. Phys.}, {\rm 48}, 113.

\bibitem[Dickel \etal  1996]{dgym96}
Dickel, J.~R., Green, A., Ye, T., \& Milne, D.~K. 1996, { \aj}, {\rm 111}, 340.

\bibitem[Dubner \etal  1996]{dgg+96}
Dubner, G.~M., Giacani, E.~B., Goss, W.~M., Moffett, D.~A., \& Holdaway, M.
  1996, { \aj}, {\rm 111}, 1304.

\bibitem[Dubner \etal  1994]{dggw94}
Dubner, G.~M., Giacani, E.~B., Goss, W.~M., \& Winkler, P.~F. 1994, { \aj},
  {\rm 108}, 207.

\bibitem[Duncan \etal  1995]{dshj95}
Duncan, A.~R., Stewart, R.~T., Haynes, R.~F., \& Jones, K.~L. 1995, { \mnras},
  {\rm 277}, 36.

\bibitem[Duric \etal  1993]{dvgv93}
Duric, N., Viallefond, F., Goss, W.~M., \& van~der Hulst, J.~M. 1993, { \aaps},
  {\rm 99}, 217.

\bibitem[Ellis \& Axon 1978]{ea78}
Ellis, R.~S. \& Axon, D.~J. 1978, { \apss}, {\rm 54}, 425.

\bibitem[Elston \& Baum 1987]{eb87}
Elston, R. \& Baum, S. 1987, { \aj}, {\rm 94}, 1633.

\bibitem[Falle \& Garlick 1982]{fg82}
Falle, S.~A. \& Garlick, A.~R. 1982, { \mnras}, {\rm 201}, 635.

\bibitem[Frail, Kassim, \& Weiler 1994]{fkw94}
Frail, D.~A., Kassim, N.~E., \& Weiler, K.~W. 1994, { \aj}, {\rm 107}, 1120.

\bibitem[Frail \& Kulkarni 1991]{fk91}
Frail, D.~A. \& Kulkarni, S.~R. 1991, { Nature}, {\rm 352}, 785.

\bibitem[Fulbright \& Reynolds 1990]{fr90}
Fulbright, M.~S. \& Reynolds, S.~P. 1990, { \apj}, {\rm 357}, 591 (FR90).

\bibitem[Gaensler \& Johnston 1995]{gj95b}
Gaensler, B.~M. \& Johnston, S. 1995, { Publ. Astron. Soc. Austral.}, {\rm 12},
  76.

\bibitem[Gaensler \etal  1997]{gms+97}
Gaensler, B.~M., Manchester, R.~N., Staveley-Smith, L., Tzioumis, A.~K.,
  Reynolds, J.~E., \& Kesteven, M.~J. 1997, { \apj}, {\rm 479}, 845.

\bibitem[Garc\'{i}a-Segura, Langer, \& Mac~Low 1996]{glm96}
Garc\'{i}a-Segura, G., Langer, N., \& Mac~Low, M.-M. 1996, { \aap}, {\rm 316},
  133.

\bibitem[Gardner \& Milne 1965]{gm65}
Gardner, F.~F. \& Milne, D.~K. 1965, { \aj}, {\rm 70}, 754.

\bibitem[Golla \& Hummel 1994]{gh94}
Golla, G. \& Hummel, E. 1994, { \aap}, {\rm 284}, 777.

\bibitem[Gray 1994]{gra94b}
Gray, A.~D. 1994, { \mnras}, {\rm 270}, 847.

\bibitem[Green 1972]{gre72}
Green, A.~J. 1972.
\newblock PhD thesis, University of Sydney.

\bibitem[Green 1984]{gre84}
Green, D.~A. 1984, { \mnras}, {\rm 209}, 449.

\bibitem[Green 1996]{gre96b}
Green, D.~A. 1996, Mullard {R}adio {A}stronomy {O}bservatory, {C}ambridge,
  {U}nited {K}ingdom (available on the {W}orld--{W}ide--{W}eb at {\it
  http://www.mras.cam.ac.uk/surveys/snrs/}).

\bibitem[Greisen 1996]{gre96c}
ed.\ E. Greisen 1996, { The AIPS COOKBOOK}, (Charlottesville: National Radio
  Astronomy Observatory).

\bibitem[Gull \& Daniell 1978]{gd78}
Gull, S.~F. \& Daniell, G.~J. 1978, { Nature}, {\rm 272}, 686.

\bibitem[Guo \& Burrows 1997]{gb97}
Guo, Z. \& Burrows, D.~N. 1997, { \apjlett}, {\rm 480}, L51.

\bibitem[Heiles 1979]{hei79}
Heiles, C. 1979, { \apj}, {\rm 229}, 533.

\bibitem[Heiles 1984]{hei84}
Heiles, C. 1984, { \apjs}, {\rm 55}, 585.

\bibitem[Huang \& Struve 1954]{hs54}
Huang, S.-S. \& Struve, O. 1954, { Annales d'Astrophysique}, {\rm 17}, 85.

\bibitem[Huang \& Thaddeus 1985]{ht85}
Huang, Y.-L. \& Thaddeus, P. 1985, { \apjlett}, {\rm 295}, L13.

\bibitem[Hummel, Beck, \& Dahlem 1991]{hbd91}
Hummel, E., Beck, R., \& Dahlem, M. 1991, { \aap}, {\rm 248}, 23.

\bibitem[Igumenshchev, Tutokov, \& Shustov 1992]{its92}
Igumenshchev, I.~V., Tutokov, A.~V., \& Shustov, B.~M. 1992, { Soviet Astron.},
  {\rm 36}, 241.

\bibitem[Insertis \& Rees 1991]{ir91}
Insertis, F.~M. \& Rees, M.~J. 1991, { \mnras}, {\rm 252}, 82.

\bibitem[Jokipii 1987]{jok87}
Jokipii, J.~R. 1987, { \apj}, {\rm 313}, 842.

\bibitem[Joncas, Roger, \& Dewdney 1989]{jrd89}
Joncas, G., Roger, R.~S., \& Dewdney, P.~E. 1989, { \aap}, {\rm 219}, 303.

\bibitem[Jun \& Norman 1996]{jn96}
Jun, B.-I. \& Norman, M.~L. 1996, { \apj}, {\rm 472}, 245.

\bibitem[Kaspi \etal  1992]{kmj+92}
Kaspi, V.~M., Manchester, R.~N., Johnston, S., Lyne, A.~G., \& D'Amico, N.
  1992, { \apjlett}, {\rm 399}, L155.

\bibitem[Kaspi \etal  1996]{kmj+96}
Kaspi, V.~M., Manchester, R.~N., Johnston, S., Lyne, A.~G., \& D'Amico, N.
  1996, { \aj}, {\rm 111}, 2028.

\bibitem[Kesteven \& Caswell 1987]{kc87}
Kesteven, M.~J. \& Caswell, J.~L. 1987, { \aap}, {\rm 183}, 118 (KC87).

\bibitem[Killeen, Bicknell, \& Ekers 1986]{kbe86}
Killeen, N. E.~B., Bicknell, G.~V., \& Ekers, R.~D. 1986, { \apj}, {\rm 302},
  306.

\bibitem[K\"{o}nigl 1982]{kon82}
K\"{o}nigl, A. 1982, { \apj}, {\rm 261}, 115.

\bibitem[Koo, Heiles, \& Reach 1992]{khr92}
Koo, B.-C., Heiles, C., \& Reach, W.~T. 1992, { \apj}, {\rm 390}, 108.

\bibitem[Koo \& Moon 1997]{km97}
Koo, B.-C. \& Moon, D.-S. 1997, { \apj}, {\rm 475}, 194.

\bibitem[Lalitha \etal  1984]{lsmt84}
Lalitha, P., Salter, C.~J., Mantovani, F., \& Tomasi, P. 1984, { \aap}, {\rm
  131}, 196.

\bibitem[Landecker \etal  1982]{lprv82}
Landecker, T.~L., Pineault, S., Routledge, D., \& Vaneldik, J.~F. 1982, {
  \apjlett}, {\rm 261}, L41.

\bibitem[Landecker \etal  1989]{lprv89}
Landecker, T.~L., Pineault, S., Routledge, D., \& Vaneldik, J.~F. 1989, {
  \mnras}, {\rm 237}, 277.

\bibitem[Large \etal  1994]{lcc+94}
Large, M.~I., Campbell-Wilson, D., Cram, L.~E., Davison, R.~G., \& Robertson,
  J.~G. 1994, { Proc. Astron. Soc. Austral.}, {\rm 11}, 44.

\bibitem[Leckband, Spangler, \& Cairns 1989]{lsc89}
Leckband, J.~A., Spangler, S.~R., \& Cairns, I.~H. 1989, { \apj}, {\rm 338},
  963.

\bibitem[Lozinskaya 1988]{loz88}
Lozinskaya, T.~A. 1988, in { Supernova remnants and the interstellar medium,
  IAU Colloquium 101}, ed.\ R.~S. Roger \& T.~L. Landecker, (Cambridge:
  Cambridge University Press), p.~95.

\bibitem[Manchester 1974]{man74}
Manchester, R.~N. 1974, { \apj}, {\rm 188}, 637.

\bibitem[Manchester 1987]{man87}
Manchester, R.~N. 1987, { \aap}, {\rm 171}, 205.

\bibitem[Manchester, D'Amico, \& Tuohy 1985]{mdt85}
Manchester, R.~N., D'Amico, N., \& Tuohy, I.~R. 1985, { \mnras}, {\rm 212},
  975.

\bibitem[Manchester \& Durdin 1983]{md83}
Manchester, R.~N. \& Durdin, J.~M. 1983, in { Supernova remnants and their
  {X}-ray emission ({IAU} {S}ymp. 101)}, ed.\ J. Danziger \& P. Gorenstein,
  (Dordrecht: Reidel), p.~421.

\bibitem[Mathewson \& Ford 1970]{mf70}
Mathewson, D.~S. \& Ford, V.~L. 1970, { Mem. R. Astron. Soc.}, {\rm 74}, 139.

\bibitem[Milne \& Haynes 1994]{mh94b}
Milne, D.~K. \& Haynes, R.~F. 1994, { \mnras}, {\rm 270}, 106.

\bibitem[Mineshige \& Shibata 1990]{ms90}
Mineshige, S. \& Shibata, K. 1990, { \apjlett}, {\rm 355}, L47.

\bibitem[Mineshige, Shibata, \& Shapiro 1993]{mss93}
Mineshige, S., Shibata, K., \& Shapiro, P.~R. 1993, { \apj}, {\rm 409}, 663.

\bibitem[Moffett \& Reynolds 1994]{mr94}
Moffett, D.~A. \& Reynolds, S.~P. 1994, { \apj}, {\rm 425}, 668.

\bibitem[Muxlow \etal  1994]{mpw+94}
Muxlow, T. W.~B., Pedlar, A., Wilkinson, P.~N., Axon, D.~J., Sanders, E.~M., \&
  de~Bruyn, A.~G. 1994, { \mnras}, {\rm 266}, 455.

\bibitem[Napier, Thompson, \& Ekers 1983]{nte83}
Napier, P.~J., Thompson, A.~R., \& Ekers, R.~D. 1983, { Proc. IEEE}, {\rm 71},
  1295.

\bibitem[Norman \& Ikeuchi 1989]{ni89}
Norman, C.~A. \& Ikeuchi, S. 1989, { \apj}, {\rm 345}, 372.

\bibitem[Norman 1993]{nor93}
Norman, M.~L. 1993, in { Back To The Galaxy: AIP Conference Proceedings 278},
  ed.\ S.~S. Holt \& F. Verter, (New York: American Institute of Physics),
  p.~552.

\bibitem[Normandeau, Taylor, \& Dewdney 1996]{ntd96}
Normandeau, M., Taylor, A.~R., \& Dewdney, P.~E. 1996, { Nature}, {\rm 380},
  687.

\bibitem[Perley 1989]{per89}
Perley, R.~A. 1989, in { Synthesis Imaging In Radio Astronomy}, ed.\ R.~A.
  Perley, F.~R. Schwab, \& A.~H. Bridle, (San Francisco: ASP Conference Series,
  Volume 6), p.~259.

\bibitem[Pineault \& Chastenay 1990]{pc90}
Pineault, S. \& Chastenay, P. 1990, { \mnras}, {\rm 246}, 169.

\bibitem[Pineault \etal  1993]{plmg93}
Pineault, S., Landecker, T.~L., Madore, B., \& Gaumont-Guay, S. 1993, { \aj},
  {\rm 105}, 1060.

\bibitem[Pineault, Landecker, \& Routledge 1987]{plr87}
Pineault, S., Landecker, T.~L., \& Routledge, D. 1987, { \apj}, {\rm 315}, 580.

\bibitem[Pineault \etal  1985]{ppl+85}
Pineault, S., Pritchet, C.~J., Landecker, T.~L., Routledge, D., \& Vaneldik,
  J.~F. 1985, { \aap}, {\rm 151}, 52.

\bibitem[Rand \& Kulkarni 1989]{rk89}
Rand, R.~J. \& Kulkarni, S.~R. 1989, { \apj}, {\rm 343}, 760.

\bibitem[Ratkiewicz, Axford, \& McKenzie 1994]{ram94}
Ratkiewicz, R., Axford, W.~I., \& McKenzie, J.~F. 1994, { \aap}, {\rm 291},
  935.

\bibitem[Reich, F\"{u}rst, \& Arnal 1992]{rfa92}
Reich, W., F\"{u}rst, E., \& Arnal, E.~M. 1992, { \aap}, {\rm 256}, 214.

\bibitem[Reid \& Silverstein 1990]{rs90}
Reid, M.~J. \& Silverstein, E.~M. 1990, { \apj}, {\rm 361}, 483.

\bibitem[Reynolds \& Fulbright 1990]{rf90}
Reynolds, S.~P. \& Fulbright, M.~S. 1990, { Proc. 21st Internat. Cosmic-Ray
  Conf. (Adelaide)}, {\rm 4}, 72.

\bibitem[Reynolds \& Gilmore 1993]{rg93}
Reynolds, S.~P. \& Gilmore, D.~M. 1993, { \aj}, {\rm 106}, 272.

\bibitem[Reynolds \& Moffett 1993]{rm93}
Reynolds, S.~P. \& Moffett, D.~A. 1993, { \aj}, {\rm 105}, 2226.

\bibitem[Roger \& Costain 1976]{rc76}
Roger, R.~S. \& Costain, C.~H. 1976, { \aap}, {\rm 51}, 151.

\bibitem[Roger \etal  1988]{rmk+88}
Roger, R.~S., Milne, D.~K., Kesteven, M.~J., Wellington, K.~J., \& Haynes,
  R.~F. 1988, { \apj}, {\rm 332}, 940.

\bibitem[R\'{o}\.{z}yczka \& Tenorio-Tagle 1995]{rt95}
R\'{o}\.{z}yczka, M. \& Tenorio-Tagle, G. 1995, { \mnras}, {\rm 274}, 1157.

\bibitem[R\'{o}\.{z}yczka \etal  1993]{rtfb93}
R\'{o}\.{z}yczka, M., Tenorio-Tagle, G., Franco, J., \& Bodenheimer, P. 1993, {
  \mnras}, {\rm 261}, 674.

\bibitem[Sault, Staveley-Smith, \& Brouw 1996]{ssb96}
Sault, R.~J., Staveley-Smith, L., \& Brouw, W.~N. 1996, { \aaps}, {\rm 120},
  375.

\bibitem[Sault, Teuben, \& Wright 1995]{stw95}
Sault, R.~J., Teuben, P.~J., \& Wright, M. C.~H. 1995, in { Astronomical {D}ata
  {A}nalysis {S}oftware and {S}ystems {IV}}, ed.\ R. Shaw, H.E. Payne, \&
  J.J.E. Hayes, ASP Conference Series, Volume 77, p.~433.

\bibitem[Seward \etal  1983]{shmc83}
Seward, F.~D., Harnden~Jr., F.~R., Murdin, P., \& Clark, D.~H. 1983, { \apj},
  {\rm 267}, 698.

\bibitem[Shaver 1969]{sha69}
Shaver, P.~A. 1969, { Observatory}, {\rm 89}, 227.

\bibitem[Shaver 1982]{sha82}
Shaver, P.~A. 1982, { \aap}, {\rm 105}, 306.

\bibitem[Shaver \etal  1985]{ssp+85}
Shaver, P.~A., Salter, C.~J., Patnaik, A.~R., van Gorkom, J.~H., \& Hunt, G.~C.
  1985, { Nature}, {\rm 313}, 113.

\bibitem[Shull, Fesen, \& Saken 1989]{sfs89}
Shull, J.~M., Fesen, R.~A., \& Saken, J.~M. 1989, { \apj}, {\rm 346}, 860.

\bibitem[Slettebak 1949]{sle49}
Slettebak, A. 1949, { \apj}, {\rm 110}, 498.

\bibitem[Sofue \& Fujimoto 1983]{sf83}
Sofue, Y. \& Fujimoto, M. 1983, { \apj}, {\rm 265}, 722.

\bibitem[Sofue, Fujimoto, \& Wielebinski 1986]{sfw86}
Sofue, Y., Fujimoto, M., \& Wielebinski, R. 1986, { \araa}, {\rm 24}, 459.

\bibitem[Stewart \etal  1993]{schn93}
Stewart, R.~T., Caswell, J.~L., Haynes, R.~F., \& Nelson, G.~J. 1993, {
  \mnras}, {\rm 261}, 593.

\bibitem[Stone \& Norman 1992]{sn92}
Stone, J.~M. \& Norman, M.~L. 1992, { \apj}, {\rm 389}, 297.

\bibitem[Storey \etal  1992]{ssmk92}
Storey, M.~C., Staveley-Smith, L., Manchester, R.~N., \& Kesteven, M.~J. 1992,
  { \aap}, {\rm 265}, 752.

\bibitem[Tenorio-Tagle, Bodenheimer, \& Yorke 1985]{tby85}
Tenorio-Tagle, G., Bodenheimer, P., \& Yorke, H.~W. 1985, { \aap}, {\rm 145},
  70.

\bibitem[Tenorio-Tagle \etal  1991]{trfb91}
Tenorio-Tagle, G., R\'{o}\.{z}yczka, M., Franco, J., \& Bodenheimer, P. 1991, {
  \mnras}, {\rm 251}, 318.

\bibitem[Tomisaka 1990]{tom90}
Tomisaka, K. 1990, { \apjlett}, {\rm 361}, L5.

\bibitem[Tomisaka 1992]{tom92}
Tomisaka, K. 1992, { Proc. Astr. Soc. Jap.}, {\rm 44}, 177.

\bibitem[Tuohy, Clark, \& Burton 1982]{tcb82}
Tuohy, I.~R., Clark, D.~H., \& Burton, W.~B. 1982, { \apjlett}, {\rm 260}, L65.

\bibitem[van~der Laan 1962]{laa62b}
van~der Laan, H. 1962, { \mnras}, {\rm 124}, 179.

\bibitem[Whiteoak \& Gardner 1968]{wg68}
Whiteoak, J.~B. \& Gardner, F.~F. 1968, { \apj}, {\rm 154}, 807.

\bibitem[Whiteoak \& Green 1996]{wg96}
Whiteoak, J. B.~Z. \& Green, A.~J. 1996, { \aaps}, {\rm 118}, 329 (WG96).

\bibitem[Williams \etal  1997]{wcd+97}
Williams, R.~M., Chu, Y.-H., Dickel, J.~R., Beyer, R., Petre, R., Smith, R.~C.,
  \& Milne, D.~K. 1997, { \apj}, {\rm 480}, 618.

\bibitem[Willingale \etal  1996]{wwps96}
Willingale, R., West, R.~G., Pye, J.~P., \& Stewart, G.~C. 1996, { \mnras},
  {\rm 278}, 749.

\bibitem[Ye, Turtle, \& Kennicutt~Jr. 1991]{ytk91}
Ye, T., Turtle, A.~J., \& Kennicutt~Jr., R.~C. 1991, { \mnras}, {\rm 249}, 722.

\bibitem[Zhang, Wang, \& Chen 1996]{zwc96}
Zhang, Q.-C., Wang, Z., \& Chen, Y. 1996, { \apj}, {\rm 466}, 808.

\end{thebibliography}

\clearpage 
\newpage

\noindent
FIGURE CAPTIONS 

\begin{figure}[ht]
\caption{
1.4 GHz VLA image of SNR \snra. The linear greyscale
is from --0.5 to 3.0 mJy beam$^{-1}$, and the peak surface brightness
in the image is 15.8 mJy beam$^{-1}$. The synthesised beam, shown 
at the lower right of the Figure, corresponds to a resolution of 
$15'' \times 9''$. The solid line represents a constant Galactic
latitude $b = -0\fdg4$. The dashed line represents the axis
of symmetry for this SNR, as discussed in the text.}
\label{fig_g003}
\end{figure}

\begin{figure}[ht]
\caption{
Image of SNR \snrb, produced by combining 1.4 GHz data
from the VLA and from Parkes. The linear greyscale
is from --2.4 to 7.5 mJy beam$^{-1}$, and the peak surface brightness
in the image is 58.5 mJy beam$^{-1}$. The synthesised beam, shown 
at the lower right of the Figure, corresponds to a resolution of 
$21'' \times 18''$. A line of constant Galactic latitude and the
axis of symmetry for the SNR are shown as in Figure~\ref{fig_g003}.  
Regions A, B and C, as discussed in the text, are indicated. 
}
\label{fig_g350}
\end{figure}

\begin{figure}[ht]
\caption{1.4 GHz image of SNR G166.0+04.3, reproduced from Pineault \etal
(1987) and precessed to J2000 coordinates. A line of
constant Galactic latitude and the axis of symmetry for the SNR are
shown as in Figure~\ref{fig_g003}.  The ``wing'' and ``shell''
components as defined by Landecker \etal (1982) are shown.}
\label{fig_vro}
\end{figure}

\begin{figure}[ht]
\caption{
0.843 GHz image of SNR G315.4--02.3 (RCW 86), reproduced from WG96.
}
\label{fig_g315}
\end{figure}

\begin{figure}[ht]
\caption{0.843 GHz image of SNR G332.4--00.4 (RCW 103), reproduced from WG96.}
\label{fig_g332.4}
\end{figure}

\begin{figure}[ht]
\caption{
1.4 GHz image of G046.8--00.3, reproduced from
Dubner \etal (1996) and precessed to J2000.  
Annotations to the image are as in Figure~\ref{fig_g003}.
}
\label{fig_g46}
\end{figure}

\begin{figure}[ht]
\caption{0.843 GHz image of SNR G296.5+10.0, reproduced from 
KC87 and precessed to J2000.
Annotations to the image are as in Figure~\ref{fig_g003}.}
\label{fig_g296}
\end{figure}

\begin{figure}[ht]
\caption{
0.843 GHz image of SNR G320.4--01.2, reproduced from 
WG96. Annotations to the image are as in Figure~\ref{fig_g003}.
}
\label{fig_g320}
\end{figure}

\begin{figure}[ht]
\caption{0.843 GHz image of SNR G327.6+14.6, reproduced from 
KC87 and precessed to J2000.
Annotations to the image are as in Figure~\ref{fig_g003}.}
\label{fig_sn1006}
\end{figure}

\begin{figure}[ht]
\caption{0.843 GHz image of SNR G332.0+00.2, reproduced from WG96. 
Annotations to the image are as in Figure~\ref{fig_g003}.}
\label{fig_g332.0}
\end{figure}

\begin{figure}[ht]
\caption{
0.843 GHz image of SNR G356.3--01.5, reproduced from Gray (1994).
Annotations to the image are as in Figure~\ref{fig_g003}.}
\label{fig_g356}
\end{figure}

\begin{figure}[ht]
\caption{
Histogram of the distribution of $\psi$, the acute angle
between the bilateral axis of a SNR and the Galactic Plane, for SNRs
listed in Table~\ref{tab_psi}.}
\label{fig_psi}
\end{figure}

\begin{figure}[ht]
\caption{
Representation of a bilateral SNR in Cartesian
space. The vector corresponds to the axis of the remnant, assuming
cylindrical symmetry.  The observed angle between the projected axis
and the Galactic Plane is  $\psi = |\phi|$.}
\label{fig_axes}
\end{figure}

\clearpage
\newpage

\begin{table}
\begin{tabular}{cccc} \tableline \tableline
Date        & Array    & Source       & Time on source \\ \tableline
1996 Jan 20 & CnB      & G003.7--00.2 & 165$^{\rm m}$ \\
	    &          & G350.0--01.8 & 90$^{\rm m}$ \\ \tableline
1996 May 21 & DnC      & G350.0--01.8 & 24$^{\rm m}$ \\ \tableline
1996 Jun 18 & DnC      & G003.7--00.2 & 31$^{\rm m}$ \\
	    &          & G350.0--01.8 & 24$^{\rm m}$ \\ \tableline
1996 Jun 22 & DnC      & G003.7--00.2 & 47$^{\rm m}$ \\
	    &          & G350.0--01.8 & 40$^{\rm m}$ \\ \tableline
\end{tabular}
\caption{VLA observations.}
\label{tab_vla}
\end{table}

\clearpage
\newpage

\begin{table}
\begin{tabular}{ll} \tableline 
Observing date     &     1996 Sep 07 \\
Centre frequency   &     1.40 GHz    \\
Bandwidth          &     40 MHz      \\
System temperature &     25 K        \\
RMS noise          &     24 mJy beam$^{-1}$ \\
Beam width (FWHM)  &   $14\farcm85 \pm 0\farcm25$ \\
Map gridding       &    $6'$ \\ \tableline
\end{tabular}
\caption{Parkes observations.}
\label{tab_pks}
\end{table}

\clearpage
\newpage

\begin{table}
\begin{tabular}{lcc} \tableline \tableline
                      &  \snra & \snrb \\ \tableline

VLA pointing centres\tablenotemark{1} &    $17^{\rm h}55^{\rm m}30^{\rm s}$, 
$-25^{\circ}51'00''$ &
(1) $17^{\rm h}28^{\rm m}30^{\rm s}$, $-38^{\circ}37'30''$ \\
 & &   (2) $17^{\rm h}27^{\rm m}20^{\rm s}$, $-38^{\circ}27'30''$ \\
 & &   (3) $17^{\rm h}26^{\rm m}55^{\rm s}$, $-38^{\circ}10'00''$ \\
 & &   (4) $17^{\rm h}26^{\rm m}00^{\rm s}$, $-38^{\circ}23'00''$ \\
Resolution & $15'' \times 9''$, PA $= 60^{\circ}$ & $21'' \times 18''$, 
PA $= -13^{\circ}$ \\
RMS noise (mJy beam$^{-1}$) & 0.06 & 0.15 \\
Derived centre of SNR\tablenotemark{1} & $17^{\rm h}55^{\rm m}28^{\rm s}$, 
$-25^{\circ}50'$ & $17^{\rm h}27^{\rm m}50^{\rm s}$, $-38^{\circ}30'$ \\
Derived centre of SNR\tablenotemark{2} & 003.78, --00.29  & 349.95, --02.03 \\
Diameter   & 13$'$ & $47' \times 44'$ \\
1.4 GHz flux density (Jy) & $1.7 \pm 0.1$ & $22.3 \pm 0.3$ \\
Spectral index & $-0.65 \pm 0.05$ & $-0.4 \pm 0.1$  \\
$S_{\rm 1\,GHz}$ (Jy) & 2.3$\pm$0.1 & 25.7$\pm$0.9 \\
$\Sigma_{\rm 1\,GHz} $ (10$^{-21}$ W m$^{-2}$ Hz$^{-1}$ sr$^{-1}$) & 
2.0$\pm$0.1 & 1.7$\pm$0.1 \\ \tableline
\end{tabular}
\tablenotetext{1}{Equatorial coordinates $\alpha$, $\delta$ (J2000)}
\tablenotetext{2}{Galactic coordinates $l$, $b$}
\caption{Observational, measured and derived parameters for each SNR.}
\label{tab_snr}
\end{table}

\begin{table}
\begin{tabular}{llccc} \tableline \tableline
SNR          & Other Name   &   $\psi$ & Figure & Reference \\ \tableline 

\snra        &              &   $6\arcdeg\pm2\arcdeg$ & \ref{fig_g003} & This paper \\
G046.8--00.3 &              &  $21\arcdeg\pm3\arcdeg$ & \ref{fig_g46} & 1 \\
G078.2+02.1 & $\gamma$ Cygni & $10\arcdeg\pm2\arcdeg$ & & 2 \\
G093.3+06.9  & DA~530       &  $15\arcdeg\pm1\arcdeg$  & & 3 \\
G127.1+00.5  &              &  $3\arcdeg\pm3\arcdeg$  & & 4 \\
G156.2+05.7  &              &  $64\arcdeg\pm4\arcdeg$ & & 5 \\
G166.0+04.3  & VRO 42.05.01 & $2\arcdeg\pm2\arcdeg$ & \ref{fig_vro} & 6, 7 \\
G296.5+10.0  & PKS 1209--51/52 & $80\arcdeg\pm3\arcdeg$ & \ref{fig_g296} & 8, 9 \\
G302.3+00.7  &              & $42\arcdeg\pm2\arcdeg$ & & 10\\
G308.8--00.1 &              &  $9\arcdeg\pm3\arcdeg$ & & 10\\
G318.2+00.1  &              & $12\arcdeg\pm3\arcdeg$ & & 10 \\
G320.4--01.2 & RCW~89, MSH 15--5{\em 2} & $7\arcdeg\pm3\arcdeg$ &\ref{fig_g320}&10\\
G327.6+14.6  & SN~1006      & $86\arcdeg\pm2\arcdeg$   & \ref{fig_sn1006} & 8, 9 \\
G332.0+00.2  &              & $25\arcdeg\pm4\arcdeg$   & \ref{fig_g332.0} &10\\
G336.7+00.5  &              & $46\arcdeg\pm1\arcdeg$   & & 10\\
\snrb        &              & $5\arcdeg\pm2\arcdeg$ & \ref{fig_g350} & This paper \\
G356.3--01.5 &             &  $2\arcdeg\pm2\arcdeg$ & \ref{fig_g356} & 11\\
\tableline 
\end{tabular}
\tablerefs{(1)~Dubner \etal (1996)\nocite{dgg+96}
(2)~Pineault \& Chastenay (1990)\nocite{pc90}
(3)~Lalitha \etal (1984)\nocite{lsmt84}
(4)~Joncas, Roger \& Dewdney (1989)\nocite{jrd89}
(5)~Reich, F\"{u}rst and Arnal (1992)\nocite{rfa92}
(6)~Landecker \etal (1982)\nocite{lprv82}
(7)~Pineault, Landecker \& Routledge (1987)\nocite{plr87}
(8)~KC87
(9)~Roger \etal (1988)\nocite{rmk+88}
(10)~WG96
(11)~Gray (1994)\nocite{gra94b}}
\caption{Bilateral SNRs and their orientation with respect to 
the Galactic Plane.}
\label{tab_psi}
\end{table}

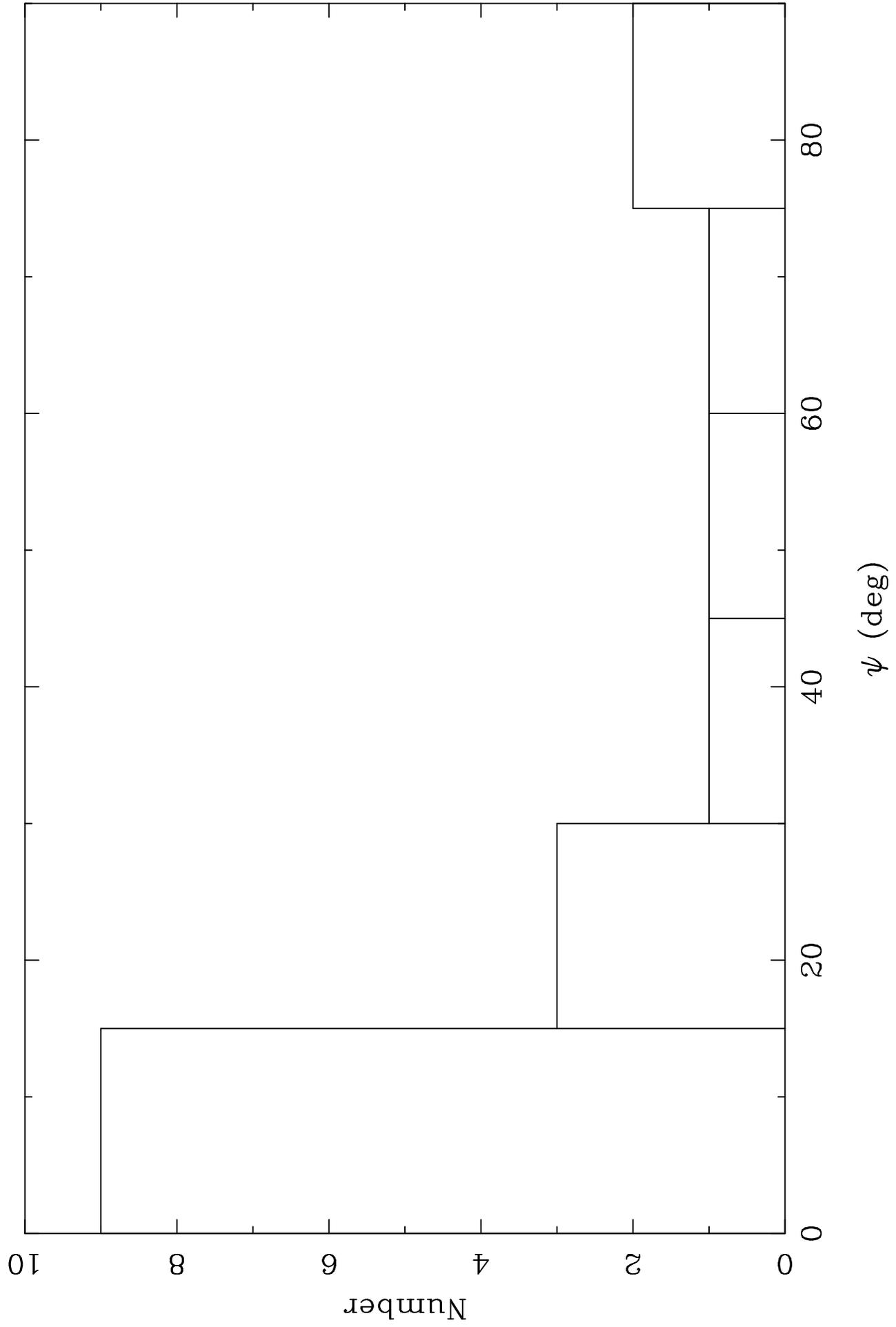
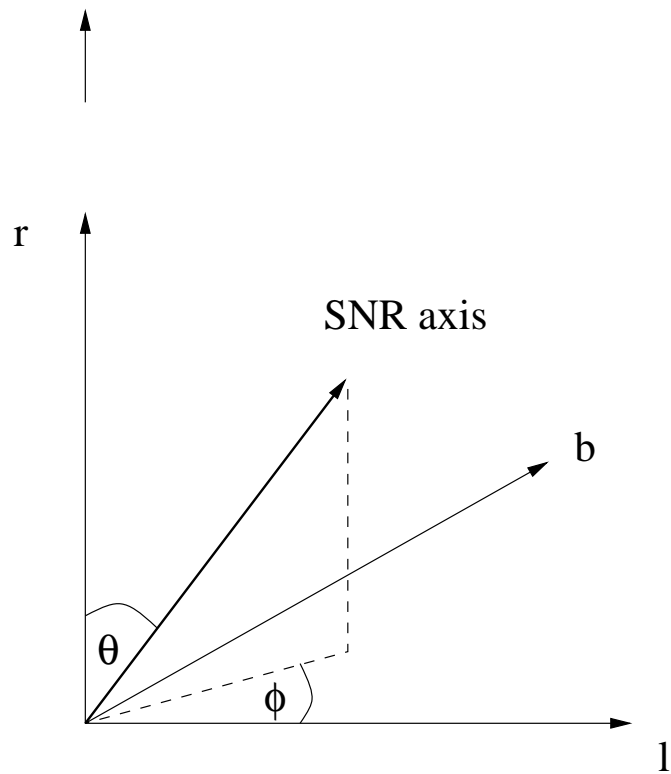
\end{document}